%% file: emnlp2020.tex
\pdfoutput=1
\documentclass[11pt,a4paper]{article}
\usepackage[hyperref]{emnlp2020}
\usepackage{times}
\usepackage{latexsym}
\usepackage{soul}
\usepackage{url}
\usepackage{graphicx}
\usepackage{amsmath}
\usepackage{booktabs}
\usepackage{latexsym}
\usepackage{microtype}
\usepackage{subfigure}
\usepackage{graphicx}
\usepackage{amssymb}
\usepackage[boxed,ruled,commentsnumbered]{algorithm2e}
\usepackage{multirow}
\usepackage{makecell}
\usepackage{verbatim}
\usepackage{changepage}

\usepackage{microtype}

\aclfinalcopy 
\def\aclpaperid{1950} 

\title{HieRec: Hierarchical User Interest Modeling for \\Personalized News Recommendation}

\author{Tao Qi$^1$, Fangzhao Wu$^2$, Chuhan Wu$^1$, Peiru Yang$^1$, Yang Yu$^2$, Xing Xie$^2$ and Yongfeng Huang$^1$\\
  $^1$Department of Electronic Engineering \& BNRist, Tsinghua University, Beijing 100084, China  \\
  $^2$Microsoft Research Asia, Beijing 100080, China\\
  {\tt \{taoqi.qt, wufangzhao, wuchuhan15, peiruyang17\}@gmail.com}\\
  {\tt  yfhuang@mail.tsinghua.edu.cn}\\
  {\tt \{t-yyu,xing.xie\}@microsoft.com} \\}

\date{}

\begin{document}
\maketitle

\begin{abstract}

User interest modeling is critical for personalized news recommendation.
Existing news recommendation methods usually learn a single user embedding for each user from their previous behaviors to represent their overall interest.
However, user interest is usually diverse and multi-grained, which is difficult to be accurately modeled by a single user embedding.
In this paper, we propose a news recommendation method with hierarchical user interest modeling, named \textit{HieRec}.
Instead of a single user embedding, in our method each user is represented in a hierarchical interest tree to better capture their diverse and multi-grained interest in news.
We use a three-level hierarchy to represent 1) overall user interest; 2) user interest in coarse-grained topics like sports; and 3) user interest in fine-grained topics like football.
Moreover, we propose a hierarchical user interest matching framework to match candidate news with different levels of user interest for more accurate user interest targeting.
Extensive experiments on two real-world datasets validate our method can effectively improve the performance of user modeling for personalized news recommendation.

\end{abstract}

\input{data/Introuduction.tex}
\input{data/Relatework.tex}

\input{data/Approach.tex}
\input{data/Experiment.tex}
\input{data/Conclusion.tex}

\section*{Acknowledgments}
This work was supported by the National Natural Science Foundation of China under Grant numbers U1936208, U1705261, U1936216, and U1836204.
We thank Tao Di and Wei He for their great comments and suggestions.

\input{data/Impact.tex}

\bibliography{anthology}
\bibliographystyle{acl_natbib}


\end{document}

%% file: data/Introuduction.tex
\section{Introduction}
Recently, massive people are habituated to reading news articles on online news platforms, such as Google News and Microsoft News~\cite{khattar2018weave,das2007google}.
To help users efficiently obtain their interested news information, personalized news recommendation technique that aims to recommend news according to user interests, is widely used by these platforms~\cite{wu2020sentirec,liu2010personalized,lin2014personalized}.

\begin{figure}
    \centering
    \resizebox{0.48\textwidth}{!}{
    \includegraphics{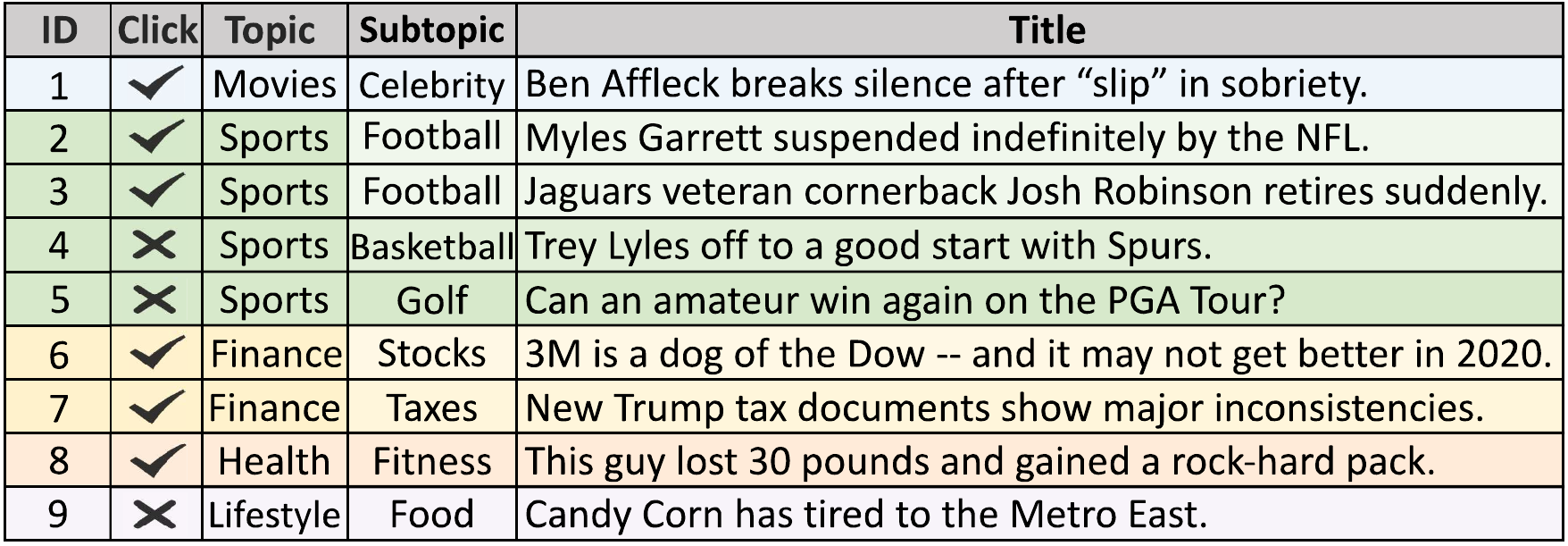}
    }
    \caption{Click and non-click logs of an example user.}
    \label{fig.intro}
    \vspace{-0.2in}
\end{figure}

User interest modeling is a critical step for personalized news recommendation~\cite{wu2020fairness,zheng2018drn,wu2020ptum}.
Existing methods usually learn a single representation vector to model overall user interests from users' clicked news~\cite{okura2017embedding,wuuser,an2019neural}.
For example, \citet{okura2017embedding} used a GRU network to model user interests from clicked news.
They used the latest hidden state of GRU as the user interest representation.
\citet{wu2019neuralc} used multi-head self-attention network to capture user interests, and used an attentive pooling network to obtain a unified user representation.
However, user interest is usually diverse and multi-grained.
For example, as shown in Fig.~\ref{fig.intro}, a user may have interest in movies, sports, finance and health at the same time.
In addition, for users who are interested in sports, some of them may have general interest in this area, while other users like the example user in Fig.~\ref{fig.intro} may only have interest in a specific sport like football.
However, it is difficult for these methods to accurately model the diverse and multi-grained user interest for news recommendation via a single user embedding.

In this paper, we propose a personalized news recommendation approach with hierarchical user interest modeling, named \textit{HieRec}, which can effectively capture the diverse and multi-grained user interest.
Our approach contains three levels of user interest representations to model user interests in different aspects and granularities.
The first one is subtopic-level, which contains multiple interest representations to model fine-grained user interests in different news subtopics (e.g., interest in football and golf).
They are learned from embeddings of subtopics and the clicked news in the corresponding subtopics.
The second one is topic-level, which contains multiple interest representations to capture coarse-grained user interests in major news topics (e.g., interest in sports and finance).
They are learned from embeddings of news topics and their subordinate subtopic-level interest representations.
The third one is user-level, which contains an interest representation to model overall user interests.
It is learned from topic-level interest representations.
Besides, we propose a hierarchical user interest matching framework to match candidate news with different levels of interest representations to target user interests more accurately.
Extensive experiments on two real-world datasets show that \textit{HieRec} can effectively improve the accuracy of user interest modeling and news recommendation.


    
    
    




%% file: data/Relatework.tex
\section{Related Work}

Personalized news recommendation is an important intelligent application and is widely studied in recent years~\cite{bansal2015content,wutanr,qi2020privacy,ge2020graph}.
Existing methods usually model news from its content, model user interest from user's clicked news, and recommend candidate news based on their relevance with user interests~\cite{okura2017embedding}.
For example, \citet{okura2017embedding} utilized an auto-encoder to learn news representations from news bodies.
They applied a GRU network to capture user interests from the sequence of users' historical clicks and used the last hidden state vector of GRU as user interest representation.
Besides, they proposed to model relevance between user interest and candidate news based on the dot product of their representations.
\citet{wu2019ijcai} learned news representations from news titles, bodies, categories, and subcategories based on an attentive multi-view learning framework.
They build user interest representation based on the attentive aggregation of clicked news representations.
\citet{an2019neural} used a CNN network to learn news representations from news titles and categories.
They applied a GRU network to user's clicked news to build a short-term user interest representation and applied user ID embedding to learn long-term user interest representation.
They further learned a unified user interest representation based on the aggregation of short- and long-term user interest representation.
\citet{liu2020kred} proposed to learn news representations from news titles and entities via a knowledge graph attention network.
They also obtained user interest representation from representations of clicked news via an attention network.
Besides, all of these three methods adopted the inner product for matching candidate news.
Most existing methods learn a single user embedding to represent the overall user interests~\cite{wang2018dkn,wu2019neuralc,wu2019npa}.
However, user interests are usually very diverse and multi-grained, which are difficult to be accurately modeled by a single user embedding.
Different from these methods, we propose a hierarchical user interest modeling framework to model user interests in different aspects and granularities.
In addition, we propose a hierarchical user interest matching framework to understand user interest in candidate news from different interest granularities for more accurate user interest targeting.

%% file: data/Approach.tex
\section{HieRec}

\begin{figure*}[!h]
    \centering
    \resizebox{0.98\textwidth}{!}{
    \includegraphics{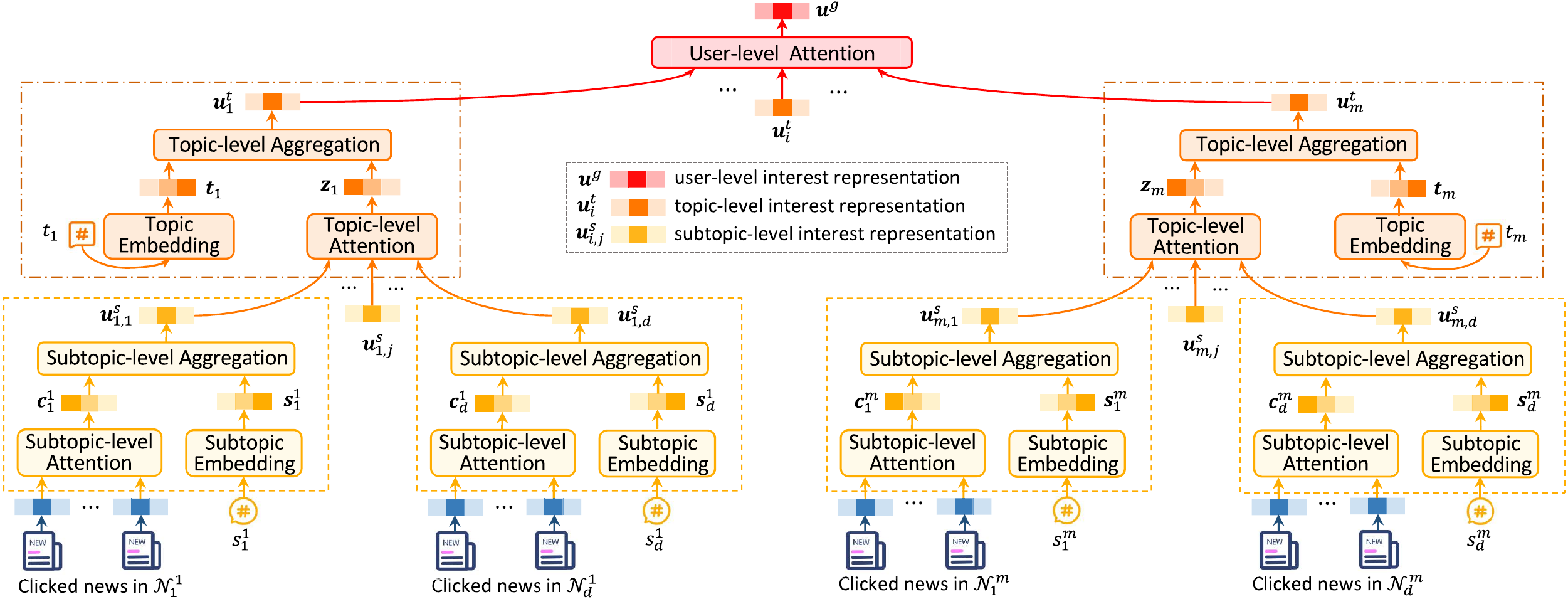}
    }
    \caption{Framework of hierarchical user interest modeling in \textit{HieRec}.}
    \label{fig.user_encoder}
\end{figure*}

In this section, we first give a problem formulation of personalized news recommendation.
Then we introduce our \textit{HieRec} method in detail.

\subsection{Problem Formulation}

Given a candidate news $n_c$ and a target user $u$, the goal is calculating an interest score $o$ to measure the interest of this user in the candidate news.
Each news $n$ has a title, a topic $t$ and a subtopic $s$.
The title is composed of a text sequence $\textbf{T} = [w_1,w2,...,w_T]$ and an entity sequence $\textbf{E} = [e_1,e_2,...,e_E]$, where $w_i$ and $e_i$ respectively denote the $i$-th word and entity in news title, $T$ and $E$ respectively denote the number of words and entities.
We assume the user has $M$ clicked news.
In \textit{HieRec}, we further divide these clicks based on their topics and subtopics for hierarchical user interest modeling.
More specifically, we build a clicked topic set $\{t_i|i=1,...,m\}$ from topics of user's clicks, where $t_i$ is the $i$-th clicked topic and $m$ is the number of clicked topics.
We can further obtain a clicked subtopic set $\{s_j^i|j=1,...,d\}$ subordinate to each clicked topic $t_i$, where $s_j^i$ is the $j$-th clicked subtopic subordinate to topic $t_i$ and $d$ is the size of the set.
Finally, user's clicked news in topic $t_i$ and subtopic $s_j^i$ are divided into the same click group $\mathcal{N}_j^i = \{n_k^{i,j}|k=1,...,l\}$, where $n_k^{i,j}$ denotes the $k$-th clicked news in this group and $l$ is the number of clicked news in the group.

\subsection{Hierarchical User Interest Modeling}

In general, user interest is usually very diverse and multi-grained.
For example, according to Fig.~\ref{fig.intro}, the example user has interests in many different aspects at the same time, such as sports, movies, and finance.
Besides, for users who are interested in sports, some of them may have general interests in this area and may read news on different kinds of sports, such as basketball, football, golf, and so on.
While other users (like the example user in Fig.~\ref{fig.intro}) may only have interest in a specific sport like football.
Understanding user interest in different aspects and granularities has the potential to model user interests more accurately.
Thus, we propose a hierarchical user interest modeling framework, which learns a hierarchical interest tree to capture diverse and multi-grained user interest.
As shown in Fig.~\ref{fig.user_encoder}, \textit{HieRec} represents user interests via a three-level hierarchy.


First, we learn multiple subtopic-level interest representations to model fine-grained user interests in different news subtopics (e.g. football and golf).
The subtopic-level interest representation for subtopic $s^i_j$ is learned from $\mathcal{N}^i_j$ that is composed of user's clicked news in subtopic $s^i_j$.
Since clicked news may have different informativeness for modeling user interest, we adopt a subtopic-level attention network to select informative clicked news for modeling user interest in subtopic $s^i_j$:
\begin{equation}
    \textbf{c}^i_j = \sum_{k=1}^l \gamma_k \textbf{n}^{i,j}_k,\  \gamma_k = \frac{ \exp(\phi_s(\textbf{n}^{i,j}_k))}{\sum_{p=1}^l \exp(\phi_s(\textbf{n}^{i,j}_p) )},
\end{equation}
where $\gamma_k$ denotes the attention weight of the $k$-th clicked news $n^{i,j}_k$ in $\mathcal{N}^i_j$, $\textbf{n}^{i,j}_k$ is the representation of news $n^{i,j}_k$ (Section.~\ref{sec.ne} introduces how to obtain it) and $\phi_s(\cdot)$ denotes a dense network.
Besides, we also adopt a subtopic embedding layer to capture semantic information of different subtopics, from which we can obtain the embedding vector $\textbf{s}^i_j$ of subtopic $s^i_j$.
Finally, we learn the subtopic-level user interest representation $\textbf{u}^s_{i,j}$ based on the combination of $\textbf{c}^i_j$ and $\textbf{s}^i_j$, i.e., $\textbf{u}^s_{i,j} = \textbf{c}^i_j + \textbf{s}^i_j$.
Similarly, we also learn subtopic-level interest representations for other subtopics clicked by the user.

\begin{figure*}[!h]
    \centering
    \resizebox{0.65\textwidth}{!}{
    \includegraphics{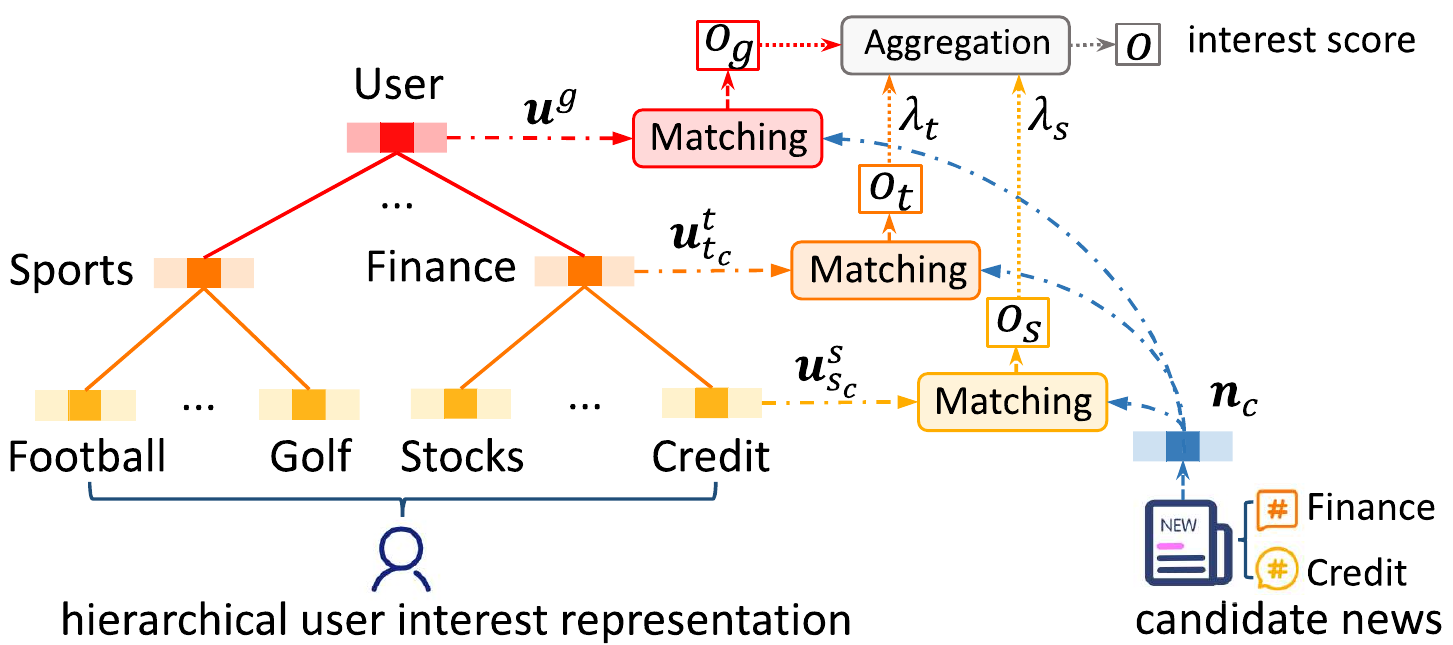}
    }
    \caption{Framework of hierarchical user interest matching in \textit{HieRec}.}
    \label{fig.matching}
\end{figure*}

Second, we learn multiple topic-level interest representations to model coarse-grained user interests in major news topics (e.g. sports and finance).
The topic-level interest representation for a clicked topic $t_i$ is learned from subtopic-level interest representations $\{\textbf{u}^s_{i,j}|j=1,...,d\}$ of subtopics $\{s_j^i|j=1,...,d\}$ subordinate to the topic $t_i$.
More specifically, user interests in different subtopics may have different importance for modeling user interest in a specific topic.
Besides, the number of clicked news on a subtopic may also reflect its importance for modeling topic-level user interest.
Thus, we utilize a topic-level attention network to select important subtopic-level user interest representations to model user interest in topic $t_i$:
\begin{equation}
    \textbf{z}_i = \sum_{j=1}^d \beta_j \textbf{u}^s_{i,j},\ \beta_j = \frac{ \exp(\phi_t(\textbf{v}^s_{i,j}))}{\sum_{k=1}^d \exp(\phi_t(\textbf{v}^s_{i,k}) )},
\end{equation}
where $\textbf{v}^s_{i,j} = [\textbf{u}^s_{i,j}; \textbf{r}^i_j]$, $\textbf{r}^i_j$ is the embedding vector for the number of clicked news on subtopic $s^i_j$, $[\cdot;\cdot]$ is the concatenation operation, $\beta_j$ is the attention weight of  $\textbf{u}^s_{i,j}$, and $\phi_t(\cdot)$ is a dense network.
Besides, we also use a topic embedding layer to model semantic information of different topics and drive the embedding vector $\textbf{t}_i$ for topic $t_i$.
Finally, we aggregate $\textbf{z}_i$ and $\textbf{t}_i$ to learn the topic-level user interest representation $\textbf{u}^t_i$ in topic $t_i$: $\textbf{u}^t_i = \textbf{z}_i + \textbf{t}_i$.
Similarly, we also learn topic-level interest representations for other clicked topics.

Third, we learn a user-level interest representation $\textbf{u}^g$ to model overall user interests.
It is learned from topic-level interest representations.
Similarly, we adopt a user-level attention network to model relative importance of topic-level user interests to learn user-level interest representation:
\begin{equation}
    \textbf{u}^g = \sum_{i=1}^m \alpha_i \textbf{u}^t_i,\ \  \alpha_i = \frac{ \exp(\phi_g(\textbf{v}^t_{i}))}{\sum_{j=1}^m \exp(\phi_g(\textbf{v}^t_{j}) )},
\end{equation}
where $\textbf{v}^t_i = [\textbf{u}^t_i; \textbf{r}_i]$, $\textbf{r}_i$ is the embedding vector for the number of user's clicked news on topic $t_i$, $\alpha_i$ denotes the attention weight of the $i$-th topic-level interest representation, and $\phi_g(\cdot)$ denotes a dense network for calculating attention scores.

\subsection{Hierarchical User Interest Matching}

Matching between candidate news and user interests at different granularities can provide various clues for user interest targeting.
For example, according to Fig.~\ref{fig.intro}, although all of the 3rd, 4th, and 5th news are about sports, the user only clicks the 3rd news probably because of her fine-grained interests in football rather than basketball and golf.
This implies that the matching between candidate news and fine-grained user interests is useful for personalized news recommendation.
Besides, not all candidate news can match with fine-grained user interests.
For instance, a news on subtopic baseball cannot match any fine-grained interests of the example user in Fig.~\ref{fig.intro}.
Fortunately, the coarse-grained user interests (i.e., interest in sports) and overall user interests can match with this candidate news.
This implies that matching candidate news with coarse-grained user interests and overall user interests is also important.
Thus, we propose a hierarchical user interest matching framework, which models user interests in candidate news from different interest granularities.
As shown in Fig.~\ref{fig.matching}, it takes candidate news (including its representation $\textbf{n}_c$, topic $t_c$ and subtopic $s_c$) and hierarchical user interest representation as input.
First, we match candidate news with overall user interests and calculate a user-level interest score $o_g$ based on the relevance between $\textbf{n}_c$ and $\textbf{u}^g$: $o_g = \textbf{n}_c \cdot \textbf{u}^g$.

Second, topic-level interest representation $\textbf{u}^t_{t_c}$ models coarse-grained user interests in the topic $t_c$ of candidate news.
It can provide coarse-grained information to understand user interest in candidate news.
Thus, we match topic-level interest representation $\textbf{u}^t_{t_c}$ with candidate news $\textbf{n}_c$ as: $\hat{o}_t =  \textbf{n}_c \cdot \textbf{u}^t_{t_c}$.
Besides, we can infer users may be more interested in topics that they have clicked more.
Thus, we weights $\hat{o}_t$ based on the ratio $w_{t_c}$ of topic $t_c$ in historical clicked news and obtained topic-level interest score $o_t$: ${o}_t = \hat{o}_t * w_{t_c}$ .
Besides, if the candidate news does not belong to any user's clicked topics, we set $o_t$ as zero directly.

Third, subtopic-level interest representation $\textbf{u}^s_{s_c}$ models fine-grained user interest in the subtopic $s_c$ of candidate news and can be used to capture fine-grained user interests in candidate news.
Thus, we match subtopic-level interest representation $\textbf{u}^s_{s_c}$ and candidate news $\textbf{n}_c$ as: $\hat{o}_s =  \textbf{n}_c \cdot \textbf{u}^s_{s_c}$
Similarly, we weights $\hat{o}_s$ based on the ratio $w_{s_c}$ of subtopic $s_c$ in user's clicked news and obtain the subtopic-level interest score: ${o}_s = \hat{o}_s * w_{s_c}$ .

Finally, interest scores of three different levels are aggregated to an overall interest score $o$:
\begin{equation}
\label{eq.score}
    o =  \lambda_s o_s + \lambda_t o_t + (1-\lambda_s-\lambda_t)o_g ,
\end{equation}
where $\lambda_t$, $\lambda_s, \in \mathbb{R}^+$ are hyper-parameters for controlling the relative importance of interest scores of different levels.
Besides, we have $\lambda_t+\lambda_s < 1$.

\subsection{News Representation}
\label{sec.ne}

\begin{figure}
    \centering
    \resizebox{0.45\textwidth}{!}{
    \includegraphics{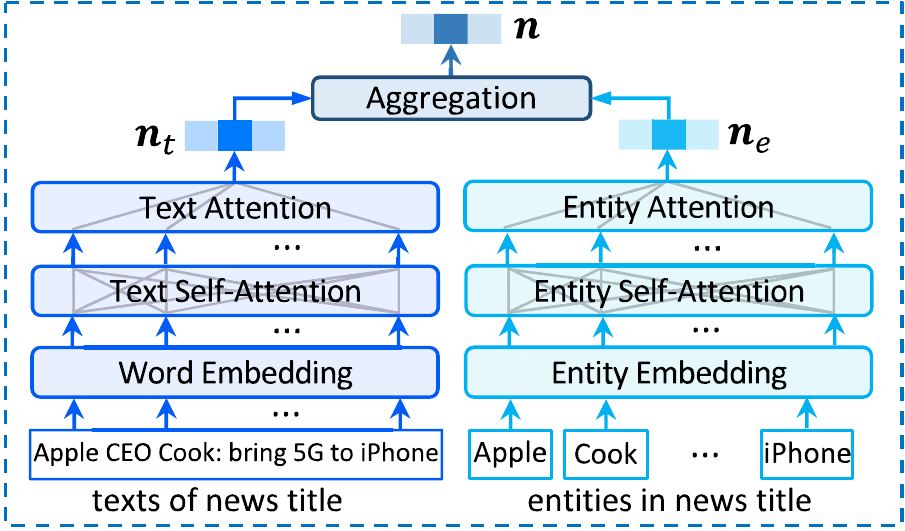}
    }
    \caption{News representation learning framework.}
    \label{fig.news_encoder}
\end{figure}

We introduce how to obtain news representation from texts and entities of news titles.
As shown in Fig.~\ref{fig.news_encoder}, we first use a text encoder to model news texts.
It first applies a word embedding layer to enrich semantic information of the model.
Next, it adopts a text self-attention network~\cite{vaswani2017attention} to learn word representations from contexts of news texts.
Then, it uses a text attention network to learn text representation $\textbf{n}_t$ by aggregating word representations.
Besides texts, knowledge graphs can also provide rich information for understanding news content via entities in news~\cite{wang2018dkn}.
Thus, we apply an entity encoder to learn entity representation of news.
We first use an entity embedding layer to incorporate information from knowledge graphs into our model.
We further apply an entity self-attention network to capture relatedness among entities.
Next, we utilize an entity attention network to learn entity representation $\textbf{n}_e$ of news by aggregating entities. 
Finally, we build representation $\textbf{n}$ of news as: $\textbf{n} = \textbf{W}_t \textbf{n}_t + \textbf{W}_e\textbf{n}_e$, where $\textbf{W}_t$ and $\textbf{W}_e$ are parameters.

\subsection{Model Training}

Following~\cite{wu2019neurald}, we utilize the NCE loss for model optimization.
Given a positive sample $n^+_i$ (a clicked news) in the training dataset $\mathcal{O}$, we randomly select $K$ negative samples $[n^1_i,...,n^K_i]$ (non-clicked news) for it from the same news impression displayed to the user $u$.
The NCE loss $\mathcal{L}$ requires the positive sample should be assigned a higher interest score $o^+_i$ than other negative samples [$o^1_i,...,o^K_i]$ and is formulated as:
\begin{equation}
  \mathcal{L} = - \sum_{i=1}^{|\mathcal{O}|} \log\frac{\exp(o^+_i)}{\exp(o^+_i)+\sum_{j=1}^K\exp(o^j_i)}.
\end{equation}

%% file: data/Experiment.tex
\section{Experiment}

\subsection{Experimental Datasets and Settings}

\begin{table}[h]
\centering
\resizebox{0.48\textwidth}{!}{
\begin{tabular}{cccccc}
\hline
      & \# News   & \# Topics & \# Subtopics & \# Users & \# Clicks \\ \hline
\textit{MIND}  & 65,238    & 18        & 270          & 94,057   & 347,727   \\
\textit{Feeds} & 1,126,508 & 28        & -            & 50,605   & 473,697   \\ \hline
\end{tabular}
}
\caption{Statistic information of the two datasets.}
\label{table.stat}
\end{table}

We conduct extensive experiments on two real-world datasets to evaluate the effectiveness of \textit{HieRec}.
The first one is the public \textit{MIND} dataset~\cite{wu2020mind}\footnote{We use the small version of MIND for quick experiments. This dataset is at https://msnews.github.io/index.html}.
It is constructed by user behavior data collected from Microsoft News from October 12 to November 22, 2019 (six weeks), where user data in the first four weeks was used to construct users' reading history, user data in the penultimate week was used for model training and user data in the last week was used for evaluation.
Besides, \textit{MIND} contains off-the-shelf topic and subtopic label for each news.
The second one (named \textit{Feeds}) is constructed by user behavior data sampled from a commercial news feeds app in Microsoft from January 23 to April 01, 2020 (13 weeks).
We randomly sample 100,000 and 10,000 impressions from the first ten weeks to construct training and validation set, and 100,000 impressions from the last three weeks to construct test data.
Since \textit{Feeds} only contains topic label of news, we implement a simplified version of \textit{HieRec} with only user- and topic- level interest representations on \textit{Feeds}.
Besides, following \citet{wu2020mind}, users in \textit{Feeds} were anonymized via hash algorithms and de-linked from the production system to protect user privacy.
Detailed information is summarized in Table~\ref{table.stat}.

\begin{table*}[]
\centering
\resizebox{0.99\textwidth}{!}{
\begin{tabular}{c|cccc|cccc}
\Xhline{1.5pt}
         & \multicolumn{4}{c|}{\textit{MIND}}                                         & \multicolumn{4}{c}{\textit{Feeds}}                                         \\ \hline
         & AUC            & MRR            & nDCG@5         & nDCG@10        & AUC            & MRR            & nDCG@5         & nDCG@10        \\ \hline
EBNR     & 61.62$\pm$0.15 & 28.07$\pm$0.18 & 30.55$\pm$0.22 & 37.07$\pm$0.21 & 63.48$\pm$0.32 & 28.01$\pm$0.18 & 32.05$\pm$0.23 & 37.64$\pm$0.22 \\
DKN      & 63.99$\pm$0.23 & 28.95$\pm$0.08 & 31.73$\pm$0.14 & 38.38$\pm$0.17 & 62.94$\pm$0.22 & 28.05$\pm$0.26 & 32.15$\pm$0.34 & 37.68$\pm$0.36 \\
DAN      & 64.68$\pm$0.13 & 29.78$\pm$0.13 & 32.63$\pm$0.21 & 39.27$\pm$0.15 & 62.67$\pm$0.49 & 27.75$\pm$0.34 & 31.74$\pm$0.44 & 37.42$\pm$0.43 \\
NAML     & 64.30$\pm$0.30 & 29.81$\pm$0.17 & 32.64$\pm$0.24 & 39.11$\pm$0.20 & 64.48$\pm$0.24 & 28.99$\pm$0.13 & 33.37$\pm$0.16 & 38.90$\pm$0.18 \\
NPA      & 64.28$\pm$0.53 & 29.64$\pm$0.33 & 32.28$\pm$0.37 & 38.93$\pm$0.39 & 64.02$\pm$0.63 & 28.71$\pm$0.39 & 33.01$\pm$0.50 & 38.55$\pm$0.47 \\
LSTUR    & 65.68$\pm$0.35 & 30.44$\pm$0.39 & 33.49$\pm$0.45 & 39.95$\pm$0.39 & 65.01$\pm$0.13 & 29.28$\pm$0.06 & 33.74$\pm$0.09 & 39.16$\pm$0.11 \\
NRMS     & 65.43$\pm$0.15 & 30.74$\pm$0.18 & 33.13$\pm$0.17 & 39.66$\pm$0.15 & 65.27$\pm$0.19 & 29.40$\pm$0.15 & 33.89$\pm$0.16 & 39.34$\pm$0.15 \\
KRED     & 65.89$\pm$0.31 & 30.80$\pm$0.32 & 33.78$\pm$0.27 & 40.23$\pm$0.26 & 65.51$\pm$0.11 & 29.57$\pm$0.06 & 34.04$\pm$0.06 & 39.60$\pm$0.05 \\
GNewsRec & 65.91$\pm$0.21 & 30.50$\pm$0.21 & 33.56$\pm$0.21 & 40.13$\pm$0.18 & 65.23$\pm$0.16 & 29.36$\pm$0.11 & 33.87$\pm$0.13 & 39.44$\pm$0.12 \\ 
FIM & 64.65$\pm$0.14 & 29.70$\pm$0.17 & 32.51$\pm$0.25 & 39.30$\pm$0.16 & 65.41$\pm$0.23 & 29.57$\pm$0.18 & 34.08$\pm$0.25 & 39.56$\pm$0.23 \\ \hline
HieRec     & \textbf{67.95}$\pm$0.14 & \textbf{32.87}$\pm$0.08 & \textbf{36.36}$\pm$0.07 & \textbf{42.53}$\pm$0.10 & \textbf{66.23}$\pm$0.10 & \textbf{29.82}$\pm$0.11 & \textbf{34.42}$\pm$0.13 & \textbf{39.94}$\pm$0.13 \\ 
\Xhline{1.5pt}
\end{tabular}
}
\caption{Performance of different methods. The improvement of \textit{HieRec} over the best baseline method is significant at level $p<0.01$ based on t-test.}
\label{table.performance}
\end{table*}

Next, we introduce experimental settings and hyper-parameters of \textit{HieRec}.
We use the first 30 words and 5 entities of news titles and users' recent 50 clicked news in experiments.
We adopt pre-trained glove~\cite{pennington2014glove} word embeddings and TransE entity embeddings~\cite{bordes2013translating} for initialization.
In \textit{HieRec}, the word and entity self-attention network output 400- and 100-dimensional vectors, respectively.
Besides, the unified news representation is 400-dimensional.
Attention networks (i.e., $\phi_s(\cdot)$, $\phi_t(\cdot)$, and $\phi_g(\cdot)$) are implemented by single-layer dense networks.
Besides, dimensions of topic and subtopic embeddings are 400, both of which are randomly initialized and fine-tuned.
The hyper-parameters for combining different interest scores, i.e. $\lambda_t$ and $\lambda_s$, are set to 0.15 and 0.7 respectively.
Moreover, we utilize dropout technique~\cite{srivastava2014dropout} and Adam optimizer~\cite{kingma2014adam} for training.
\textit{HieRec} is trained for 5 epochs with 0.0001 learning rate.
All hyper-parameters of \textit{HieRec} and baseline methods are manually tuned on the validation set.\footnote{ https://github.com/JulySinceAndrew/HieRec}
Following~\citet{wu2019neuralc}, we use four ranking metrics, i.e., AUC, MRR, nDCG@5, and nDCG@10, for performance evaluation.

\subsection{Main Results}

We first introduce the baseline methods we compared in experiments:
(1) \textit{EBNR}~\cite{okura2017embedding}: learning user representations from the sequence user's clicked news via a GRU network.
(2) \textit{DKN}~\cite{wang2018dkn}: using a candidate-aware attention network to learn user representations.
(3) \textit{DAN}~\cite{danzhu2019}: using an attentive LSTM network to learn user representations.
(4) \textit{NAML}~\cite{wu2019ijcai}: learning user representations by attentively aggregating user's clicked news.
(5) \textit{NPA}~\cite{wu2019npa}: learning news and user representations via personalized attention networks.
(6) \textit{LSTUR}~\cite{an2019neural}: modeling short-term user interests from user's clicked news via a GRU network and long-term user interests from user-news interactions via user ID embeddings.
(7) \textit{NRMS}~\cite{wu2019neuralc}: applying multi-head self-attention networks to learn news representations and user representations.
(8) \textit{KRED}~\cite{liu2020kred}: proposing a knowledge graph attention network to learn news representations from texts and entities of news titles.
(9) \textit{GNewsRec}~\cite{hu2020graph}: modeling short-term user interests from clicked news sequences via an attentive GRU network and long-term user interests from user-news click graph via a graph neural network.
(10) \textit{FIM}~\cite{wang2020fine}: modeling user interests in candidate news from semantic relevance of user's clicked news and candidate news via a 3-D CNN network.

Each experiment is repeated 5 times.
The average results and standard deviations are listed in Table~\ref{table.performance}, from which we have several observations.
First, \textit{HieRec} significantly outperforms other baseline methods which learn a single user embedding to model overall user interests, such as \textit{NRMS}, \textit{NPA}, and \textit{NAML}.
This is because user interests are usually diverse and multi-grained.
However, it is difficult for a single representation vector to model user interests in different aspects and granularities, which may be suboptimal for personalized news recommendation.
Different from these methods, we propose a hierarchical user interest modeling framework, which can represent diverse and multi-grained user interests via a three-level hierarchy.
Besides, we also propose a hierarchical user interest matching framework to match user interest with candidate news from different granularities, which can better target user interests.
Second, \textit{HieRec} can significantly outperform \textit{FIM}, which directly model user interests in candidate news from the semantic relevance of candidate news and user's clicked news.
This may be because \textit{FIM} did not consider user interests from different granularities for matching candidate news.

\subsection{Effectiveness in User Modeling}

To fairly compare different methods with \textit{HieRec} on the performance of interest modeling, we compare them based on the same news modeling method (the news modeling method introduced in Section~\ref{sec.ne}).
Experimental results are summarized in Table~\ref{table.ue} and we only show experimental results on \textit{MIND} in the following sections. 
Table~\ref{table.ue} shows that \textit{HieRec} significantly outperforms existing interest modeling methods.
This is because user interests are usually diverse and multi-grained.
It is difficult for existing methods with single user embedding to capture user interests in different aspects and granularities.
Different from these methods, \textit{HieRec} learns a three-level hierarchy to represent diverse and multi-grained user interests.


\begin{table}[]
\centering
\resizebox{0.5\textwidth}{!}{
\begin{tabular}{ccccc}
\Xhline{1.5pt}
  & AUC            & MRR            & nDCG@5         & nDCG@10        \\ \hline
NAML      & 65.81$\pm$0.27          & 30.89$\pm$0.21          & 34.16$\pm$0.30          & 40.55$\pm$0.24          \\ 
DKN & 66.03$\pm$0.27          & 31.17$\pm$0.25          & 34.47$\pm$0.33          & 40.85$\pm$0.29          \\ 
EBNR     & 65.90$\pm$0.27       & 30.86$\pm$0.21         & 34.14$\pm$0.30          & 40.58$\pm$0.24          \\
LSTUR     & 66.02$\pm$0.14       & 31.16$\pm$0.15         & 34.37$\pm$0.15          & 40.83$\pm$0.12          \\
GNewsRec  & 66.16$\pm$0.14         & 31.19$\pm$0.05         & 34.40$\pm$0.09          & 40.82$\pm$0.10         \\
NRMS      & 66.04$\pm$0.21          & 31.20$\pm$0.19          & 34.53$\pm$0.22          & 40.89$\pm$0.18          \\
\hline
HieRec      & \textbf{67.95}$\pm$0.14 & \textbf{32.87}$\pm$0.08 & \textbf{36.36}$\pm$0.07 & \textbf{42.53}$\pm$0.10 \\ 
\Xhline{1.5pt}
\end{tabular}
}
\caption{Effect of \textit{HieRec} in user interest modeling.}
\label{table.ue}
\end{table}

\begin{figure}
    \centering
    \resizebox{0.48\textwidth}{!}{
    \includegraphics{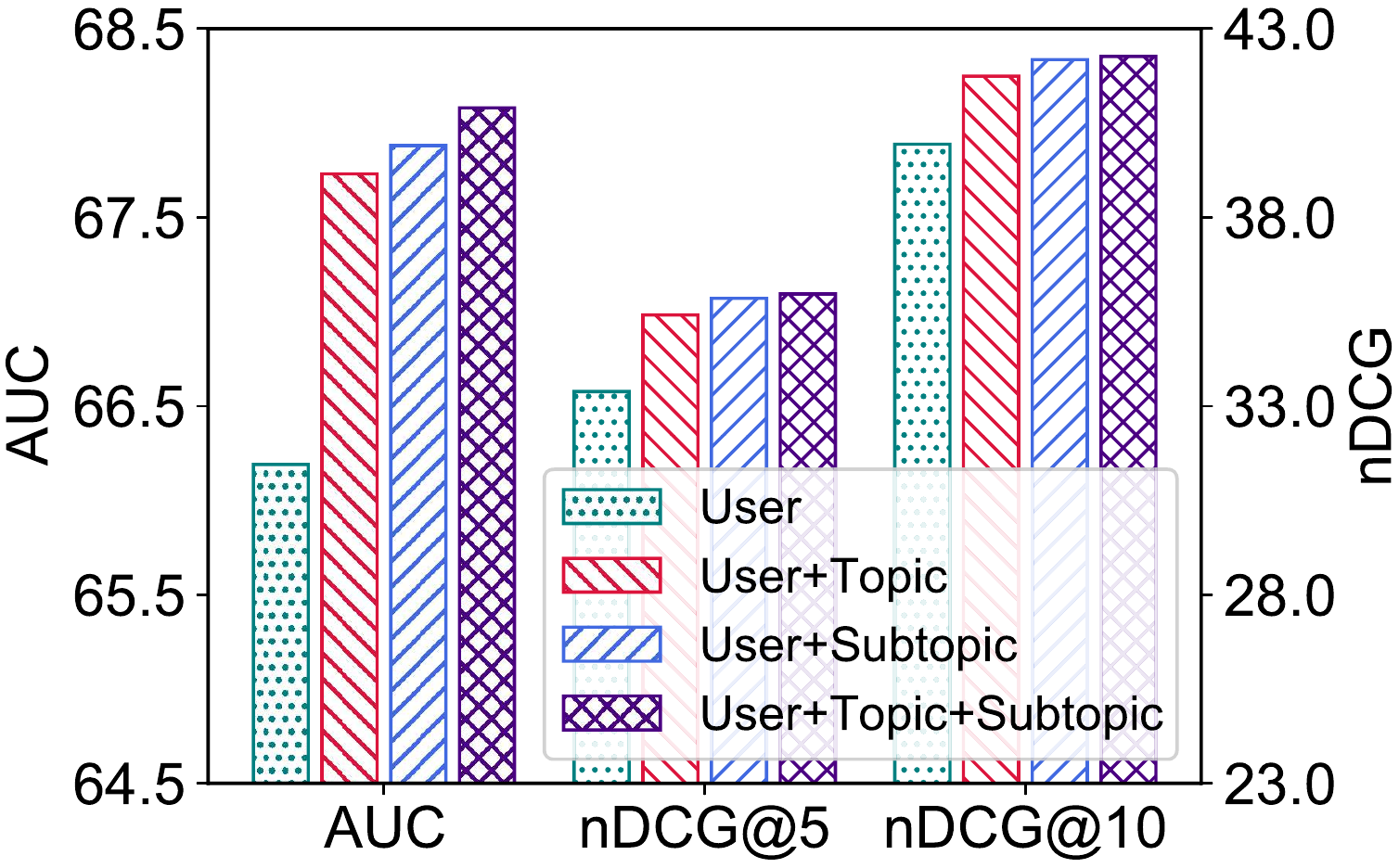}
    }
    \caption{Effect of hierarchical user interest modeling.}
    \label{fig.ablation}
\end{figure}

\subsection{Ablation Study}

We evaluate the effectiveness of user interest representations of different levels by removing the corresponding interest matching scores from Eq.~\ref{eq.score}.
Results are shown in Fig.~\ref{fig.ablation} and we have several findings.
First, \textit{HieRec} with user- and topic- or subtopic-level interest representation significantly outperforms \textit{HieRec} with only user-level interest representation.
This is because matching candidate news with fine-grained user interests has the potential to improve the accuracy of news recommendation. 
Topic- and subtopic-level interest representation can model finer-grained user interests than the user-level interest representation.
Thus, they can provide additional information to match candidate news than user-level interest representation.
Second, \textit{HieRec} with interest representations of three levels also outperforms \textit{HieRec} with user- and topic- or subtopic-level interest representation.
This may be because matching candidate news with user interests of different granularities can help perform more accurate interest matching.
Since topic- and subtopic-level interest representation capture user interests at different granularities, incorporating both of them can further improve the recommendation performance.

\subsection{Performance on Recall and Diversity}


\begin{figure}
    \centering
    \resizebox{0.37\textwidth}{!}{
    \includegraphics[clip]{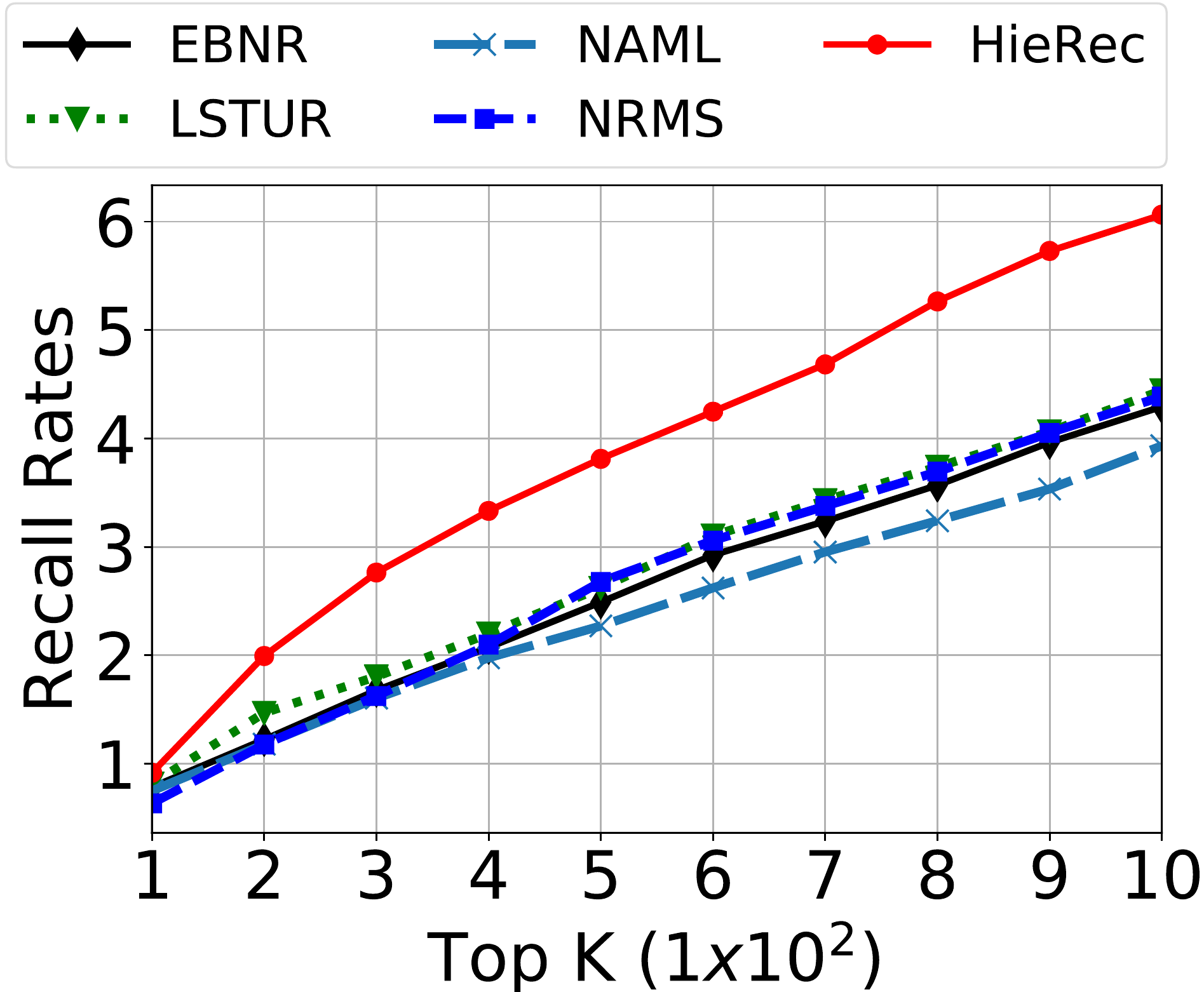}
    }
    \caption{Recall rates of different methods.}
    \label{fig.recall}
\end{figure}

\begin{figure}
    \centering
    \resizebox{0.39\textwidth}{!}{
    \includegraphics[clip]{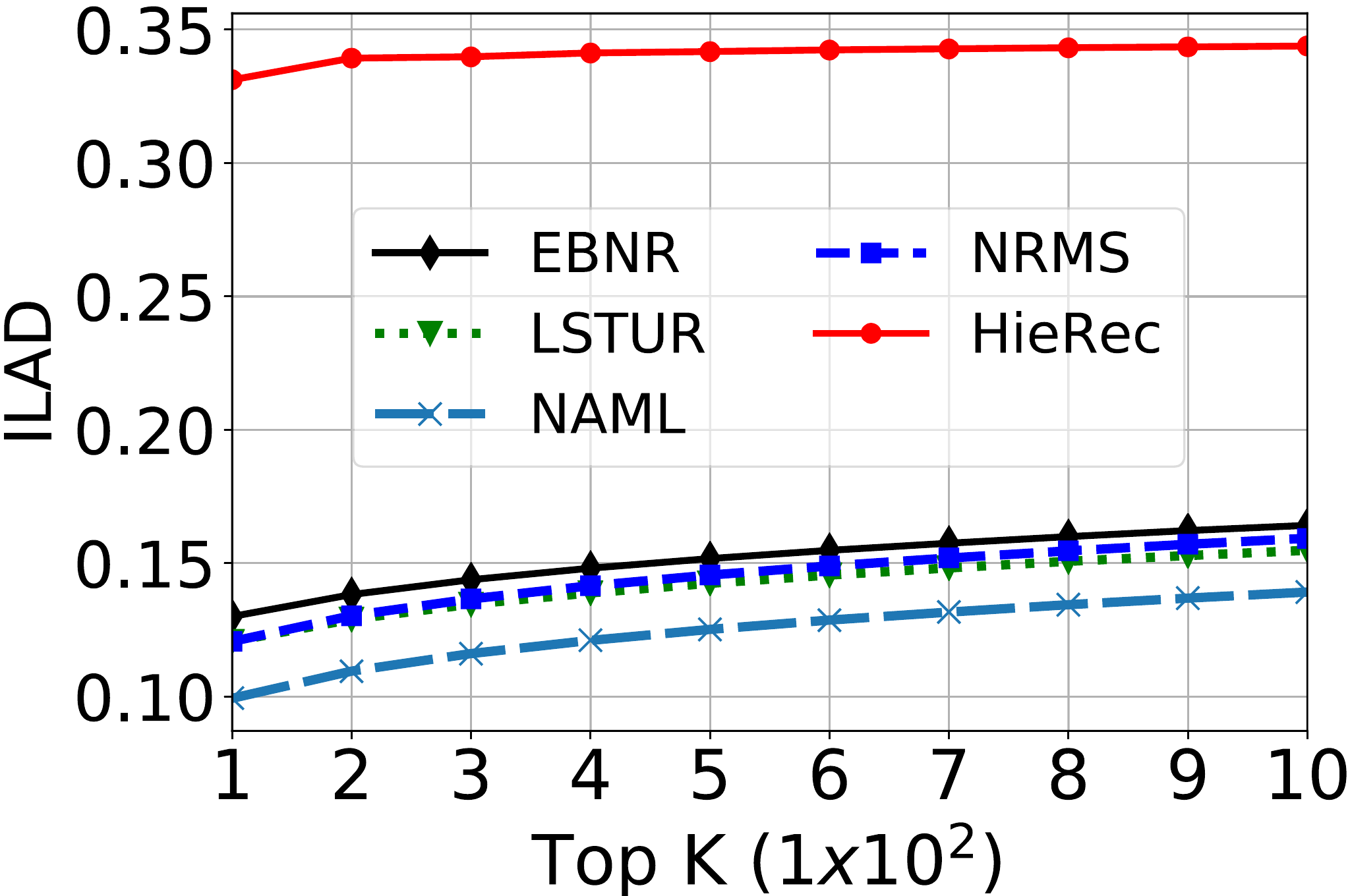}
    }
    \caption{Performance of different methods on recommendation diversity.}
    \label{fig.ilad}
\end{figure}

\begin{figure*}
    \centering
    \resizebox{0.99\textwidth}{!}{
    \includegraphics{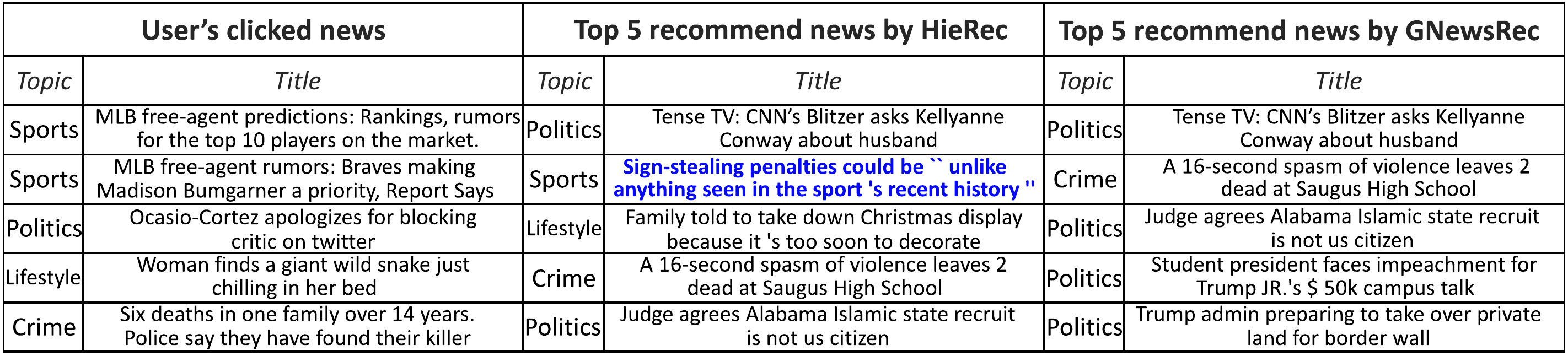}
    }
    \caption{Recommendation results of \textit{HieRec} and \textit{GNewsRec} for the same impression. The news clicked by the user in this impression is in blue and bold.}
    \label{fig.case}
\end{figure*}

Next, we compare different user interest modeling methods on the news recall task.\footnote{News recall task aims to recall a small number of candidate news from a large news pool according to user interests.}
Since methods that model user interests with candidate news information, e.g., \textit{DKN} and \textit{GNewsRec}, cannot be applied in the news recall task due to efficiency issues~\cite{pal2020pinnersage}, we do not compare them in experiments.
We evaluate the accuracy and diversity of top $K$ recalled candidate news.
Following existing works~\cite{pal2020pinnersage,chen2018fast}, the former is measured by recall rates, and the latter is measured by intra-list average distance (ILAD).
For \textit{HieRec}, we employ subtopic-level interest representations to perform multi-channel news recall and equally integrate news recalled by different interest channels.
Experimental results are summarized in Fig.~\ref{fig.recall} and Fig.~\ref{fig.ilad}, which show that \textit{HieRec} significantly outperforms other methods in terms of both recall rates and diversity.
This is because user interests are usually very diverse and multi-grained, which are difficult to be comprehensively modeled by a single representation vector.
Different from these methods, \textit{HieRec} hierarchically represents user interests and can better model user interests in different aspects and granularities.
Besides, this also implies that compared to existing personalized methods, \textit{HieRec} can help users explore more diverse information and alleviate filter bubble issues~\cite{nguyen2014exploring} to some extent.

\begin{figure}
    \centering
    \resizebox{0.5\textwidth}{!}{
    \includegraphics{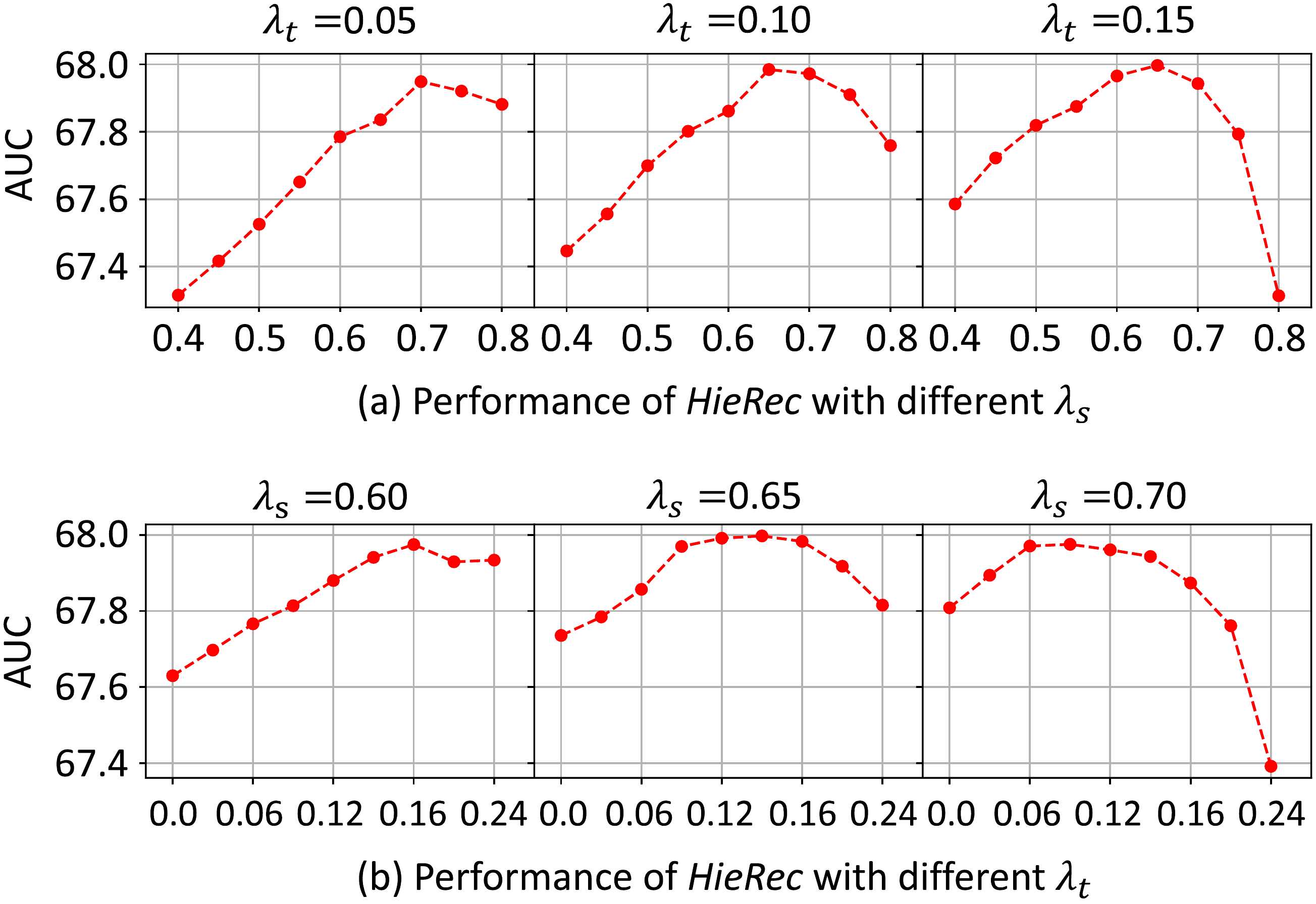}
    }
    \caption{Influence of hyper-parameters.}
    \label{fig.hyper}
\end{figure}

\subsection{Hyper-parameters Analysis}

As shown in Fig.~\ref{fig.hyper}, we analyze the influence of two important hyper-parameters of \textit{HieRec} (i.e., $\lambda_t$, $\lambda_s$) used for combining different levels of interest scores.
First, when $\lambda_t$ is fixed, performance of \textit{HieRec} first gets better with the increase of $\lambda_s$.
This is because $\lambda_s$ controls the importance of $o_s$.
Bedsides, $o_s$ measures the relevance of candidate news and fine-grained user interests, which can provide accurate information to understand user interests in the candidate news.
When $\lambda_s$ is too small, \textit{HieRec} cannot effectively exploit information in $o_s$.
Second, large value of $\lambda_s$ also hurts the performance of \textit{HieRec}.
This is because when $\lambda_s$ is too large, \textit{HieRec} cannot effectively exploit user- and topic-level matching scores to recommend candidate news.
However, matching candidate news with both overall and coarse-grained user interests is important for personalized news recommendation.
Thus, a moderate $\lambda_s$, i.e., 0.65 or 0.7, is suitable for \textit{HieRec}.
Third, when $\lambda_s$ is fixed, the performance of \textit{HieRec} also first gets better with the increase of $\lambda_t$ and gets worse when $\lambda_t$ is too large.
This is because \textit{HieRec} cannot effectively utilize information of $o_t$ when $\lambda_t$ is too small.
Besides, \textit{HieRec} cannot effectively utilize information of $o_g$ and $o_s$ when $\lambda_t$ is too large.
Thus, a moderate $\lambda_t$, i.e., 0.12 or 0.15, is suitable for \textit{HieRec}.

\subsection{Case Study}

We conduct a case study to show the superior performance of \textit{HieRec}.
We compare \textit{HieRec} with \textit{GNewsRec} since \textit{GNewsRec} achieves best AUC score in Table~\ref{table.performance} among baseline methods.
In Fig.~\ref{fig.case}, we show the top 5 news recommended by \textit{HieRec} and \textit{GNewsRec} in a randomly sampled impression.
Besides, we also show the historical clicks of the target user in this impression.
We can find that the top 5 news recommended by \textit{GNewsRec} is dominated by news on politics, which cannot comprehensively cover different user interests.
This is because user interests are usually diverse and multi-grained.
However, it is difficult for \textit{GNewsRec}, which learns a single representation to model overall user interests, to effectively capture user interests in different aspects and granularities.
Different from \textit{GNewsRec}, the top 5 news recommended by \textit{HieRec} are diverse and can cover topics that the user may be interested in.
Besides, the user clicked a news recommended by \textit{HieRec}.
This is because \textit{HieRec} learns a hierarchical user interest representation which can effectively model user interests in different aspects and granularities.
With the help of the hierarchical user interest representation, \textit{HieRec} can match candidate news with user interests in different aspects and granularities.

%% file: data/Conclusion.tex
\section{Conclusion}

In this paper, we propose a personalized news recommendation method named \textit{HieRec} for hierarchical user interest modeling, which can effectively model diverse and multi-grained user interests.
\textit{HieRec} learns a three-level hierarchy to represent user interest in different aspects and granularity.
First, we learn multiple subtopic-level interest representations to model fine-grained user interests in different news subtopics.
Second, we learn multiple topic-level interest representations to model coarse-grained user interests in several major news topics.
Third, we learn a user-level interest representation to model overall user interests.
Besides, we propose a hierarchical user interest matching framework to match candidate news with user interest from different granularity for more accurate user interest targeting.
Extensive experiments on two real-world datasets show the effectiveness of \textit{HieRec} in user interest modeling.

%% file: data/Impact.tex
\section*{Ethics and Impact Statement}

In this paper, we present \textit{HieRec} to model diverse and multi-grained user interest.
\textit{HieRec} can be applied to online news platforms for personalized news recommendation, which can help platforms improve user experience and help users find interested news information.
Although \textit{HieRec} can bring many benefits, it may also have several potential risks, which we will discuss in detail.

\textbf{Accuracy} Although \textit{HieRec} outperforms baseline methods in term of recommendation accuracy (Table~\ref{table.performance}), it may also have some inaccurate recommendation results that users are not interested in.
Users usually just ignore them and will not click them to read.
The user experience may be harmed and users may use the online news service less in the future, or turn to other online news platforms.

\textbf{Privacy}
In \textit{HieRec}, we rely on user behavior data centrally stored on the news platform for model training and online services.
User behavior data is usually privacy-sensitive, and its centralized storage may lead to privacy concerns and risks.
In the future, we will explore to train and deploy \textit{HieRec} in a more privacy-preserving way based on some effective privacy protection techniques like Federated Learning~\cite{qi2020privacy}.

\textbf{Diversity}
Filter bubbles and echo chambers are the common problem for many recommender systems~\cite{nguyen2014exploring}, which harms user experience.
Improving recommendation diversity has the potential to alleviate the problem of filter bubbles and echo chambers.
Through experiments in Fig.~\ref{fig.ilad}, we find that \textit{HieRec} can outperform many news recommendation methods in term of recommendation diversity.
Thus, compared with existing methods, \textit{HieRec} has the potential to alleviate filter bubble problem to some extent.
Besides, in order to further improve recommendation diversity, \textit{HieRec} can be combined with some existing methods in this field like DPP~\cite{chen2018fast}.

\textbf{Fake News and Clickbait}
There may be some fake news and clickbait in some online platforms.
In order to handle the negative social impact and the user experience harm brought by these fake news and clickbait, online news platforms can use some existing fake news detection and clickbait detection techniques such as~\cite{kumar2018identifying,shu2019defend} to filter these kinds of news before applying \textit{HieRec} for personalized recommendation.

\textbf{Fairness}
Like many other recommender systems, \textit{HieRec} relies on user behavior data for model training and online service.
The bias in user behavior data may lead to some specific groups of users not be able to receive news information with sufficient accuracy and diversity, and the recommendation results may be more suitable for some major populations.
Recently, some fairness-aware recommendation methods like FairRec~\cite{wu2020fairness} have been proposed to eliminate bias and unfairness in recommender systems.
We can combine \textit{HieRec} with these methods to improve the fairness of the recommendation results and mitigate the harms for marginalized populations.

\textbf{Misuse}
The proposed \textit{HieRec} method works in a data-driven way.
It trains the model from the user logs and makes personalized recommendations to users based on their interest inferred from their clicked news.
However, in some extreme cases, the recommendation results may be maliciously manipulated to influence users.
To avoid the potential misuse, the usage of \textit{HieRec} should comply with the regulations and laws, and intentional manipulation should be prohibited.